\documentclass[12pt]{iopart}
\usepackage{graphicx}
\expandafter\let\csname equation*\endcsname\relax
\expandafter\let\csname endequation*\endcsname\relax 
\usepackage{amsmath}
\usepackage{amsfonts}
\usepackage{color}
\usepackage{hyperref}

\DeclareMathOperator{\sech}{sech}

\newcommand{\Fig}{figure}

\begin{document}

\title[High-$m$ Kink/Tearing modes]{High-$m$ Kink/Tearing Modes in Cylindrical Geometry}

\author{J~W~Connor$^{1,2,3}$, R~J~Hastie$^{1,2}$, I~Pusztai$^{4,5}$,
  P~J~Catto$^{2,4}$ and M~Barnes$^{1,2,6}$}

\address{$^1$ CCFE, Culham Science Centre, Abingdon, Oxon, OX14 3DB, UK}  
\address{$^2$ Rudolf Peierls Centre for Theoretical Physics, 1 Keble Rd,
Oxford, OX1 3NP, UK}
\address{$^3$ Imperial College, London, SW7 2BW, UK}
\address{$^4$ Plasma Science and Fusion Center, MIT, Cambridge, MA 02139, USA}
\address{$^5$ Applied Physics, Chalmers University of Technology, G\"oteborg,
SE-41296, Sweden}
\address{$^6$ Department of Physics, University of Texas at Austin, Austin,
TX 78712, USA}
\ead{Jack.Connor@ccfe.ac.uk}
\begin{abstract}
 The global ideal kink equation, for cylindrical geometry and zero
 beta, is simplified in the high poloidal mode number limit and
 used to determine the tearing stability parameter,
 $\Delta^\prime$. In the presence of a steep monotonic current gradient,
 $\Delta^\prime$ becomes a function of a parameter, $\sigma_0$,
 characterising the ratio of the maximum current gradient to magnetic
 shear, and $x_s$, characterising the separation of the resonant
 surface from the maximum of the current gradient. In equilibria
 containing a current "spike", so that there is a non-monotonic
 current profile, $\Delta^\prime$ also depends on two parameters:
 $\kappa$, related to the ratio of the curvature of the current
 density at its maximum to the magnetic shear, and $x_s$, which now
 represents the separation of the resonance from the point of maximum
 current density. The relation of our results to earlier studies of
 tearing modes and to recent gyro-kinetic calculations of current
 driven instabilities, is discussed, together with potential
 implications for the stability of the tokamak pedestal.
\end{abstract}

\maketitle

\section{Introduction}
\label{sec1}

Most studies of micro-instabilities consider those driven by gradients
of density or ion and electron temperatures. However the toroidal
current gradient is also a potential source of instability. Indeed,
this is the instability drive for tearing modes and it was even
proposed \cite{KP} that it could drive high mode number ideal MHD kink
instability. This analysis was based on the periodic, cylindrical,
ideal MHD equation for the perturbed magnetic flux function,
$\Psi=\psi(r)\exp[i(m\theta-nz/R)]$, in the tokamak limit,
$B_\theta/B_z \sim r/R \ll 1$:
\begin{equation}
 \frac{d}{dr}r\frac{d \psi}{dr}-\frac{m^2}{r}\psi-\frac{
   \hat{J}^\prime }{1/q-n/m}\psi=0,
\label{eq1}
\end{equation}
 where $r\,(0<r<a)$ is the radial coordinate with
 $^\prime$ denoting a radial derivative, $\theta$ the azimuthal angle
 and $z$ the axial co-ordinate, $2\pi R$ being the periodicity length
 in the $z$-direction, and we have introduced 'poloidal' and
 'toroidal' mode numbers, $m$ and $n$, respectively. The magnetic
 field in the $z$-direction is $B_0$, the safety factor $q=rB_0/R
 B_\theta$ and the normalised current density, $\hat{J}$, is defined
 by
 \begin{equation}
 \hat{J}(r)=\frac{4\pi}{c}\frac{R J(r)}{B_0}.
\label{eq2}
 \end{equation}
 
 Expanding the denominator of (\ref{eq1}), $(1/q-n/m)$, in the
 vicinity of the resonant position, $r_s$, where, $m-nq(r_s)=0$, so
 that
 \begin{equation}
 \frac{1}{q}-\frac{n}{m}\simeq -\frac{s x}{mq},
\label{eq3}
 \end{equation}
 with
 \begin{equation}
  x = m(r-r_s)/r_s,
\label{eq4}
 \end{equation}
  denoting a new (dimensionless) local radial variable, this equation
  takes the local form:
 \begin{equation}
 \frac{d^2\psi}{dx^2}-\left(1+\frac{\sigma_0}{x}\right)\psi=0,
 \label{eq5}
 \end{equation}
 where 
 \begin{equation}
 \sigma_0=-\frac{r_s \hat{J}^\prime}{ns},
 \label{eq6}
 \end{equation}
 with $s=rq^\prime/q$ being the magnetic shear at $r=r_s$.
 Equation~(\ref{eq5}) has been the basis for a number of studies at
 high $m$. Thus Kadomtsev and Pogutse~\cite{KP} used it to claim that
 ideal MHD instability is possible, for steep current gradients or low
 shear, if
 \begin{equation}
 \sigma_0 > 2,
 \label{eq7} 
 \end{equation}
 while Wesson~\cite{JAW} and Strauss~\cite{Hank} investigated the
 dependence of the tearing mode index, $\Delta^\prime$~\cite{FKR} on
 $\sigma_0$. Hegna and Callen~\cite{HC} used a generalisation of
 (\ref{eq5}) that took account of the effect of toroidal and
 plasma shaping on metric coefficients, to estimate $\Delta'$ in more
 general, toroidal, devices. Furthermore, the local gyrokinetic code
 GS2 has recently~\cite{PIF} been used to study current gradient
 driven instabilities in collisionless plasmas, again finding
 instability above a critical value of $\sigma_0$. However this code
 is based on the ballooning transformation and this also relies on
 a linear expansion of $q$ about the rational surface.
 
Unfortunately such treatments are not entirely consistent because, for
$\sigma_0 \sim \mathcal{O}(1)$, the truncated expansion of $(q-m/n)$ employed is
inadequate. To see this is the case recall $\hat{J}(r)$ and
$q(r)$ are related through Amp$\grave{e}$re's equation,
 \begin{equation}
 \hat{J}=\frac{1}{r}\frac{d}{dr}\left(\frac{r^2}{q}\right),
 \label{eq8}
 \end{equation}
 which can be written $\hat{J}=(2-s)/q$, so that
 \begin{equation}
  r\hat{J}^\prime=(2s^2-3s-w)/q,
 \label{eq9}
 \end{equation}
 with $w$ denoting the quantity $r^2q^{\prime \prime}/q$.  Now if we
 restrict attention to positive values of the current density,
 $\hat{J}(r)$, then $s<2$ and $ | 2s^2-3s | \sim \mathcal{O}(1)$.  The
 condition $ | r \hat{J}^\prime| \gg 1$ then implies $| w | \gg 1$ so
 that terms in $q^{\prime \prime}$ must be retained in the expansion
 of $q$ around $r_s$, giving $\sigma_0=w/(m s)$.  Consequently, if the
 parameter $\sigma_0$ is of order unity, as in \cite{KP,JAW,Hank}, the
 tearing equation must now take the form:
 \begin{equation}
 \frac{d^2\psi}{dx^2}-\left[ 1+\frac{\sigma_0}{x(1+\sigma_{0}
     x/2)}\right] \psi=0.
 \label{eq10}
 \end{equation} 
 Equation~(\ref{eq10}), however, of necessity describes a scenario in
 which there are two resonant surfaces; a situation which cannot be
 the case for a monotonic $q(r)$ even when $\sigma_0\sim
 \mathcal{O}(1)$. We conclude that both (\ref{eq5}) and (\ref{eq10})
 must give a flawed description of stability of monotonic $q(r)$
 profiles. Clearly, higher order terms in the expansion of $q-n/m$ are
 required. Thus, remarkably, there is no truly local theory for such
 high-$m$ instabilities. We note that a comparison of analytic
 resistive tearing mode growth rates with those obtained from a
 numerical code at low$-m$ also required the inclusion of more
 derivatives of $q(r)$ to obtain agreement when the resistivity and
 $\Delta^\prime$ were relatively high \cite{Fulvio}.
 
 In this paper we reconsider the solution of (\ref{eq1}) at high $m$,
 taking full account of the structure of $\hat{J}(r)$ and $q(r)$. Two
 scenarios are studied: (i) a monotonic $\hat{J}(r)$ with a steep
 gradient, modelled by a "tanh function"; and (ii) the effect of a
 positive current "spike", possibly arising in the pedestal region of
 a tokamak in H-mode due to bootstrap currents. This latter case can
 lead to a region of greatly reduced shear, or even non-monotonic
 $q(r)$ and multiple resonances.  The outcomes of our calculations are
 self-consistent forms for $\Delta^\prime$ which can be used to
 investigate high-$m$ tearing mode instability. A value of
 $\Delta^\prime$ that accounts for key nonlocal features of the
 global structure of the self-consistent $q$ and current profiles is
 required to properly interpret the GS2 results alluded to above. The
 quantity $\Delta^\prime$ also plays a role in determining the
 saturated amplitude of magnetic islands arising from tearing mode
 instability. However other parameters may play a role in determining
 this amplitude \cite{poye}.
                  
\section{Reduction of the tearing equation for high-\texorpdfstring{$m$}{m}: Monotonic current profiles.}
\label{sec2}
      
\begin{figure}
\centering \includegraphics[width=8cm]{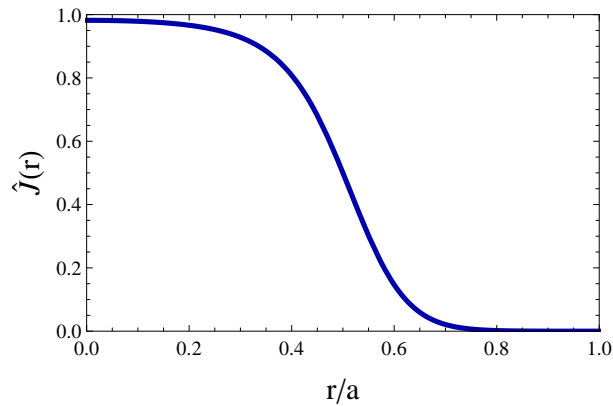}
\caption{$\hat{J}(r)$ of (\ref{eq11}), with $J_0=1$, $\lambda=4$ and
$r_0/a=0.5$.}
\label{fig1}
\end{figure}

In this section we consider a specific example for the current density
$\hat{J}(r)$ which has a steep gradient, and exploit the
property $m \gg 1$ to expand (\ref{eq1}) close to the resonant
surface at $r_s$. The current density employed is
\begin{equation}
 \hat{J}(r)=\frac{J_0}{2} \left\lbrace 1 - \frac{\tanh\lbrack
   \lambda(r^2-r_0^2)/(2 r_0^2)\rbrack}{\tanh \lbrack
   \lambda(a^2-r_0^2)/(2r_0^2)\rbrack}\right\rbrace,
\label{eq11}
\end{equation}
where $r_0$ is the point of steepest current gradient and we will be
interested in large values of the parameter $\lambda \sim
\mathcal{O}(m)$. The current profile of (\ref{eq11}) is shown in
\Fig~\ref{fig1} for $J_0=1$, $\lambda=4$ and $r_0/a=0.5$.
Integrating the Amp$\grave{e}$re equation, (\ref{eq8}), yields an
expression for the resonant denominator in the tearing equation,
(\ref{eq1}), viz.:
\begin{align}
 \frac{1}{q(r)}-& \frac{n}{m} =  \frac{J_0 r_0^2}{2 \lambda t}
 \left\lbrace\frac{1}{r_s^2} \log\left\lbrack \cosh
 \frac{\lambda(r_s^2-r_0^2)}{2r_0^2}\right\rbrack\right.  \nonumber \\ -&\frac{1}{r^2}\log
 \left\lbrack \cosh \frac{\lambda(r^2-r_0^2)}{2r_0^2}\right\rbrack
  +
 \left.\log\left(\cosh\frac{\lambda}{2}\right) \left(
 \frac{1}{r^2}-\frac{1}{r_s^2}\right) \right\},\label{eq12}\\ t \equiv
 &\tanh\left[\frac{\lambda}{2 r_0^2}(a^2-r_0^2)\right],\nonumber
\end{align}
where $r_s$ denotes the location of the resonance, i.e.
\begin{equation}
 q(r_s)=\frac{m}{n}.
\label{eq13}
\end{equation}
 The numerator of the resonant term in (\ref{eq1}) is readily
 evaluated to give
\begin{equation}
r\hat{J}^\prime= - \frac{J_0 \lambda r^2}{2 t r_0^2} \sech^2 
 \left[ \frac{\lambda}{2 r_0^2}(r^2-r_0^2)\right].
\label{eq14}
\end{equation}

\begin{figure}    
\centering\includegraphics[width=8cm]{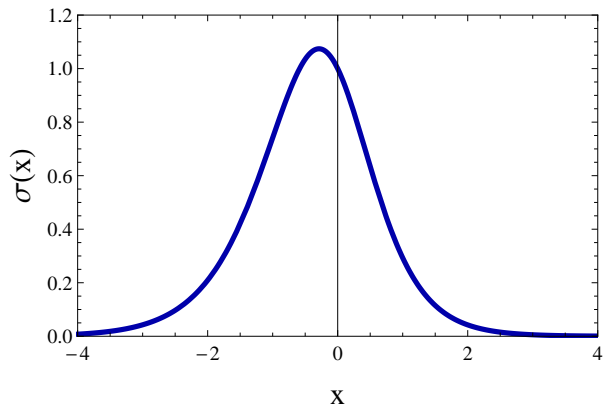}
\caption{Plot of $\sigma(x)$ for the $\hat{J}$ of \Fig~\ref{fig1},
  $p=\sigma(0)=1$ and $x_s=0$.}
\label{fig2}
\end{figure}

We next introduce the local radial variable $x=m(r-r_0)/r_0$, where we
have now chosen $r_0$ rather than $r_s$ as origin, and corresponding
resonance position, $x_s=m(r_s-r_0)/r_0$, and express $\hat{J}(x)$ and
$q(x)$ as functions of $x$ around $r=r_0$, noting that these
expressions are not confined to small values of $x$, since we only
require $ |x/m| \ll 1$. Therefore, using $t=1$ and ignoring
$\mathcal{O}(1/m^2)$ corrections in $(r_s/r_0)^2=1+(2x_s/m)$, we find
\begin{align}
\hat{J}(x) =&\frac{J_0}{2}\left\lbrack
1-\tanh(px)\right\rbrack,\label{eq15}\\ 
\frac{1}{q(x)}-\frac{1}{q(x_s)}=&-\frac{J_0}{2\lambda}
\left\{ p(x-x_s)+\log[\cosh(px)] -\log[\cosh(p x_s)]
\right\},\label{eq16}\\ 
r\hat{J}^\prime=&-\frac{\lambda
  J_0}{2}\sech^2(px),\label{eq17}\\ 
p\equiv&\frac{\lambda}{m},\label{eq18}
\end{align}
where integrating (\ref{eq8}) from $r = 0$ to $r_0$ gives $J_0
q(r_0) = 2 +\mathcal{O}(1/\lambda)$ for $\lambda \sim m \gg 1$.  When
$x_s=0$, we may write the kink/tearing equation in the form of
(\ref{eq5}), with
\begin{equation}
   \sigma_0\rightarrow \sigma(x)\equiv \frac{p^2 x
     \sech^2(px)}{px+\log[\cosh(px)]}.
\label{eq19}
\end{equation}
Since we can identify $p=\sigma(0)$ there is again a single parameter,
$\sigma(0)$, determining stability, with
\begin{equation}
   \sigma(x)=\sigma(0) f[\sigma(0) x],\qquad f (y)=\frac{y
     \sech^2(y)}{y+\log[\cosh(y)]}.
\label{eq20}
\end{equation}
Thus we can consider (\ref{eq19}) as a semi-localized
generalisation of (\ref{eq6}) in which $\sigma_0$ acquires a form
factor, $f$.  The function $\sigma(x)$ is shown in \Fig~\ref{fig2} for
$\sigma(0)=1$, from which it is clear that $\sigma(x)$ is very far
from being constant, as assumed in the local treatments of \cite{KP}
and \cite{JAW}.

Furthermore, if we assume that the mode numbers, $m$ and $n$ are such
that the resonant surface lies at $x=x_s$, i.e. that $m/n=q(x_s)$
rather than $m/n=q(0)$, the tearing equation can be written in what we
will refer to as the semi-local form:
\begin{equation}
\frac{d^2 \psi}{dx^2}-\psi\left\{ 1+\frac{p^2 \sech^2(p
  x)}{p(x-x_s)+\log[\cosh(px)]-\log[\cosh(p x_s)]}\right\}=0,
\label{eq21}
\end{equation}
 with dependence on two parameters, $p\equiv \sigma(0)$ and $x_s$. 
        
\section{Calculation of the tearing mode index, \texorpdfstring{$\Delta^\prime$}{Delta-prime}.}
\label{sec3}

In this section we consider reconnecting instabilities which might be
driven by the energy source associated with a steep monotonic current
gradient, i.e. instabilities which are driven by positive values of
the tearing index, $\Delta^\prime$, defined as the jump in
  $\psi'/\psi$ as in \cite{FKR}. However we first discuss the ideal
MHD stability properties of (\ref{eq21}), noting that these
can be determined by computing the stability properties of the two
Newcomb \cite{newcomb} sub-intervals, $[0,r_s]$ and $[r_s,a]$,
or, in terms of the "local" variable, $x$, $[-\infty,x_s]$ and
$[x_s,+\infty]$.  This analysis has been carried out numerically and no
instability found at any finite values of the parameters $p$ and $x_s$
(i.e., in Newcomb terms, in shooting a solution which is regular at
one end-point of a sub-interval, no zero of $\psi(r)$ is encountered
before the other end of the sub-interval is reached).  In particular,
the value $p=\sigma(0)=2$ is NOT a marginal point for ideal MHD
instability, contrary to Ref.~\cite{KP}.
      
Returning to the issue of tearing mode stability, we calculate
$\Delta(p,x_s)$, defined as the jump in $\psi^\prime(x)/\psi(x)$ at
$x_s$ for the solution of (\ref{eq21}) which vanishes at $x \rightarrow
-\infty$ and at $x \rightarrow +\infty$.  The global value of
$\Delta^\prime$ is then related to $\Delta$ by:
\begin{equation}
   r_s  \Delta^\prime=m \Delta.
\label{eq22}
\end{equation}

\begin{figure}    
\centering\includegraphics[width=8cm]{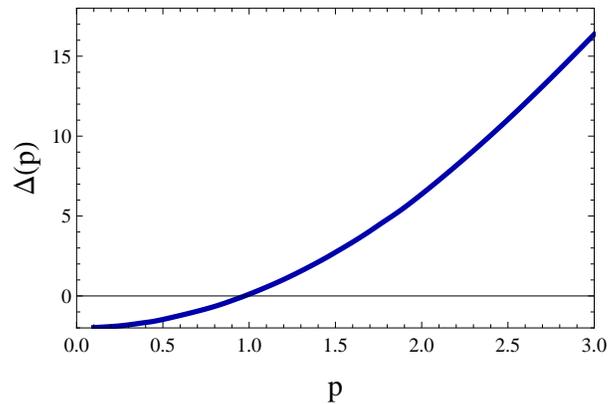}
\caption{$\Delta(p)$, for $x_s=0$ as a function of $p=\sigma(0)$
  calculated from (\ref{eq21}). The transition to
  unstable tearing occurs at $p\approx 0.97$}
\label{fig3}
\end{figure}
\begin{figure}  
\centering\includegraphics[width=8cm]{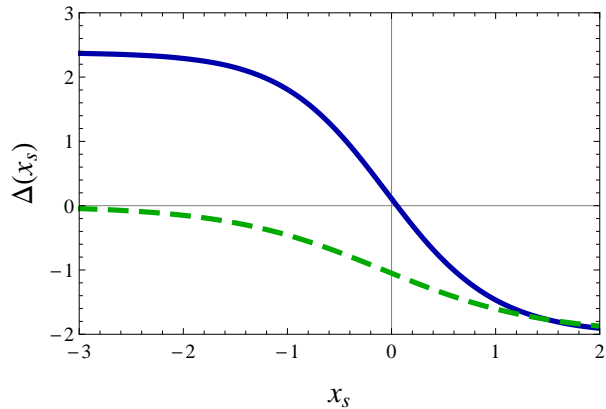}
\caption{Plot of $\Delta(x_s)$, calculated from (\ref{eq21}) for
  $p=1$ (solid curve) and $p=0.67$ (dashed curve).}
\label{fig4}
\end{figure}
  
Figure 3 shows $\Delta$ as a function of $p$ when $x_s=0$, and
\Fig~\ref{fig4} shows $\Delta$ as a function of $x_s$ for $p=1$
(solid line) and $p=0.67$ (dashed line).  The somewhat surprising
content of \Fig~\ref{fig4}, namely that the most unstable location for
the resonant surface is at large negative values of $x_s$, is an
artefact of our semi-local approximation, as discussed below.
 
\begin{figure}   
\centering\includegraphics[width=8cm]{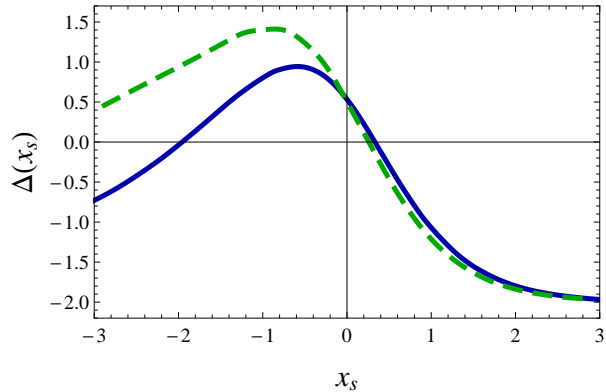} 
\caption{ Plot of $\Delta(x_s)$ as a function of $x_s$, similar to
  \Fig~\ref{fig4} but calculated from solution of the global kink/tearing
  equation, (\ref{eq1}), with $\hat{J}(r)$ given by
  (\ref{eq11}). $r_0/a=0.5$ and $m=\lambda=8$ (solid curve) and
  $\lambda=m=12$, (dashed curve), parameters which correspond to $p=1$
  in the local approximation. As described in the text,
  (\ref{eq22}) has been used to present the results in terms of
  the "local" variables, $\Delta$ and $x_s$.}
\label{fig5}
\end{figure}  
   
We have also performed global calculations of $\Delta=[r_s
  \Delta'(r_s)]/m$, with $\hat{J}(r)$ given by (\ref{eq11}).  The
results of such global calculations are shown in \Fig~\ref{fig5}
where, for the purpose of comparing with \Fig~\ref{fig4}, (\ref{eq22})
has been used to transform "global" data into "local" variables, as in
\Fig~\ref{fig4}.  In \Fig~\ref{fig5} parameters are $r_0/a=0.5$ and
$\lambda=m=8$, (solid curve), $\lambda=m=12$, (dashed curve). Figure 5
shows that the most unstable value of $\Delta$ occurs at a finite
negative value of $x_s$.  A more careful inspection of the derivation
of (\ref{eq21}) reveals that, at large negative values of $ x_s$ the
shear, $s(x_s)$, becomes exponentially small, $\sim \exp(-p x_s)$, so
that terms of order $1/m$, which have been neglected, can compete,
leading to the behaviour seen in \Fig~\ref{fig5}, for large, but
finite, $m$.  Not surprisingly, the $\Delta'$ values obtained, and
therefore stability, are sensitive to the global structure of the
ideal MHD region since equation (\ref{eq21}) contains less global
information than equation (\ref{eq11}), thereby resulting in the
$\Delta$ differences shown in figures~\ref{fig4} and \ref{fig5}.

\section{Comparison with previous studies.}
\label{sec4}
     
For comparison with the previous results of Kadomtsev and Pogutse
\cite{KP} and Wesson \cite{JAW}, we have also calculated the tearing
stability index $\Delta_0(p)$ obtained by computing solutions of
(\ref{eq5}) with $\sigma_0=p$.
\begin{figure}    
\centering\includegraphics[width=8cm]{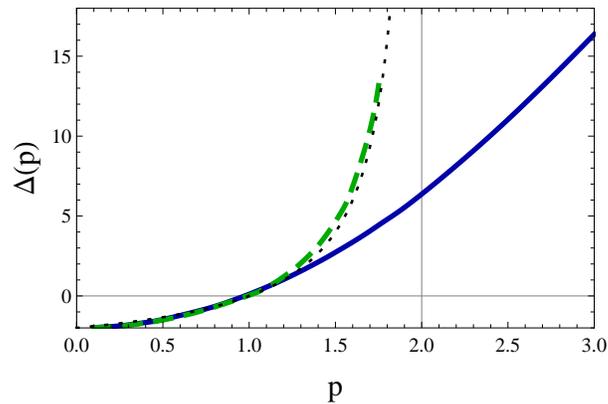}
\caption{Comparison of $\Delta_0$ with \Fig~\ref{fig3}. The solid
  curve was computed with $\sigma(x)$ given by (\ref{eq19}), the
  dashed curve was computed with constant $\sigma=p$. $\Delta_0$
  passes through zero at $p=1$, as found in \cite{JAW} and
  asymptotes to $\infty$ at $p=2$ as in \cite{KP}. The thin
  dotted curve represents a rational approximation to the dashed
  curve, used in section~\ref{sec6}.}
\label{fig6}
\end{figure}
 
Figure~\ref{fig6} shows (dashed curve) the resulting $\Delta_0(p)$ as
a function of $p$ from (\ref{eq5}). For comparison the solid
line shows $\Delta(p)$ of \Fig~\ref{fig3}, computed with
$\sigma=\sigma(x)$ of (\ref{eq19}). Ideal MHD marginality,
where the value of $\Delta_0(p)$ becomes infinite, apparently occurs
at $p=2$ which corresponds to the prediction~\cite{KP} of ideal kink
instability beyond this value. At the value $p=2$, an exact solution
of (\ref{eq5}) in the inner Newcomb sub-interval is $\psi=x e^x$,
which vanishes at both end-points, again demonstrating ideal
marginality. However, as noted in the foregoing discussion, this is an
incorrect prediction and is not found in full cylindrical solutions,
or in our semi-localised version (\ref{eq19}) with the correct
$\sigma(x)$ dependence, which yields the solid curve in
\Fig~\ref{fig6}. For small values of $p$, $\Delta_0$ is negative and
changes sign at $p\approx 1$, as reported in \cite{JAW}.

\section{Reduction of the tearing equation for high-\texorpdfstring{$m$}{m}:  non-monotonic current profiles.}
\label{sec5}  
  
\begin{figure}  
\centering\includegraphics[width=8cm]{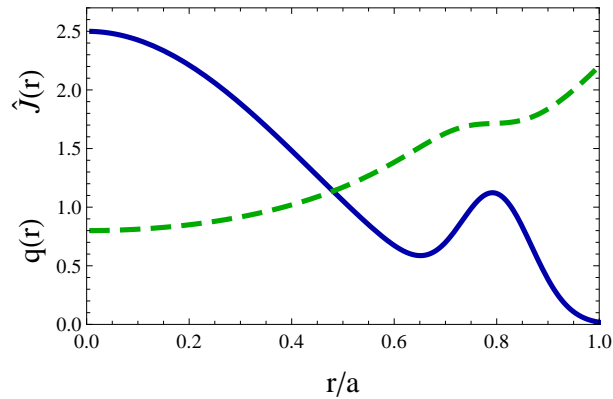}
\caption{Example of the non-monotonic class of current profile
  investigated in section~\ref{sec6}. The dashed curve shows the $q(r)$
  profile displaying reduced magnetic shear in the vicinity of the
  current spike, $J_1(r)$. Parameters are $J_0=2.5,~ J_1=1,~ \mu=64$
  and $r_1/a=0.8$.}
\label{fig7}
\end{figure}

In this section we investigate a different, but possibly important,
scenario in which a fairly localised positive current spike occurs
relatively near the plasma edge \cite{PJM, Mast}, where resistivity is
high and the inductive current density is small. The non-linear theory
of external kink modes in the presence of such a current spike has
been studied by Eriksson and Wahlberg \cite{eriksson}.  Bootstrap
currents in the region of a steep pedestal, or localized current
drive, may produce just such a situation. Figure~\ref{fig7} shows an
example for the current density given by,
\begin{eqnarray}
  \hat{J}(r)&=&\hat{J}_0(r)+\hat{J}_1(r),\nonumber \\
               &=&J_0[1-(r/a)^2]^3+J_1 e^{-\mu (r/r_1-1)^2},
\label{eq23}
\end{eqnarray}
with $J_0=2.5$, $J_1=1$, $\mu=64$ and the peak of the current spike at
$r_1/a=0.8$. In \Fig~\ref{fig7} the dashed curve represents the resulting safety
factor, $q(r)$, indicating that the magnetic shear becomes small in
the region of the current spike. 

\begin{figure}
\centering\includegraphics[width=8cm]{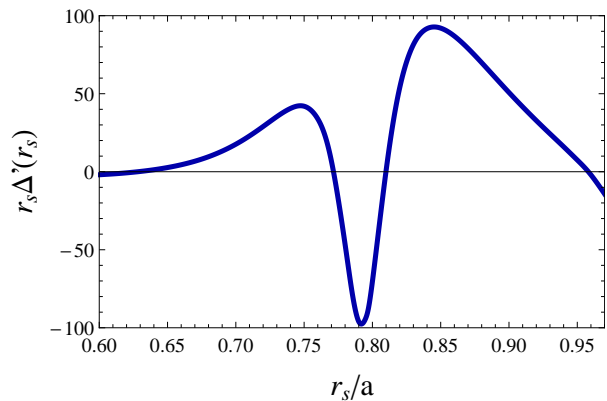}
\caption{Plot of $r_s \Delta^\prime(r_s)$ for an $m=4$ mode, from
  solution of (\ref{eq1}) with $\hat{J}(r)$ given by
  (\ref{eq23}).  }
\label{fig8}
\end{figure}

It might be thought that the most unstable location for a tearing mode
resonance would be at the minimum of the current density ($r/a=0.65$
in \Fig~\ref{fig7}), since the gradient of the current density is
destabilising on both sides of the resonance in this case, that is,
$J^\prime / [(1/q) - 1/q(r_s)] < 0$ so its sign is the opposite of the
stabilizing $m^2$ line bending term in (\ref{eq1}). Surprisingly,
however, this is not the case, as can be seen from \Fig~\ref{fig8}
which displays the value of $r_s \Delta^\prime(r_s)$ for an $m=4$
mode, calculated from (\ref{eq1}), as $r_s$ is moved across the
$\hat{J}(r)$ profile, regarding $n$ as a continuous variable.

Figure~\ref{fig8} does show that when $r_s$ is located at the maximum
of the current spike, $r_s/a\approx 0.79$ in \Fig~\ref{fig7}, the
tearing index, $\Delta^\prime$ is strongly stabilising (i.e. negative)
as one would expect since the current gradients on both sides of $r_s$
are stabilising in this case, that is, $J^\prime/[(1/q) - 1/q(r_s)] >
0$, enhancing the stabilizing $m^2$ line bending term in
(\ref{eq1}). However, values of $r_s$ quite close to the maximum of
the current spike are very unstable.
   
\begin{figure}  
\centering\includegraphics[width=8cm]{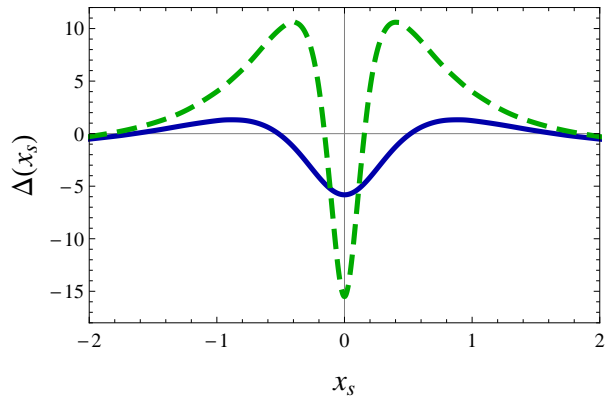}
\caption{ Plot of $\Delta(x_s)$, from (\ref{eq22}) with
  $\kappa=+8$ (solid curve) and $\kappa=64$ (dashed curve),
  corresponding to monotonically increasing $q$ profiles.}
\label{fig9}
\end{figure}

\begin{figure}
\centering\includegraphics[width=8cm]{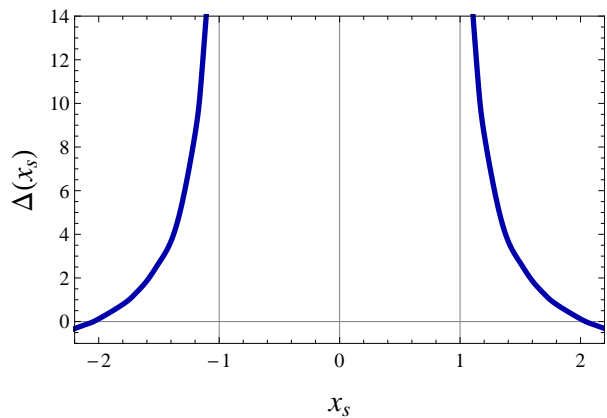}
\caption{Equivalent plot to \Fig~\ref{fig9} but with $\kappa=-8$ corresponding
  to reverse shear at $x=0$, i.e to a locally non-monotonic $q$
  profile. $\Delta$ values are only calculated when there is a single
  resonant surface, i.e. in the range $|x_s|>1$.}
\label{fig10}
\end{figure}

To model this in a high-$m$, localised analysis, we expand
$\hat{J}(r)$ and $q(r)$ locally around the maximum of the current
density, approximately at $r_1$, the maximum of the current
spike. Around this point, $\hat{J}(r)$ is dominated by the current
spike, $\hat{J}_1(r)$, and is a local function, $\hat{J}_1(x)$, if we
order the parameter $\mu \sim \mathcal{O}(m)$.  However, the $q$ profile in this
region is determined by both the extended "inductive" current profile
$\hat{J}_0(r)$ and by the current spike. It therefore contains both a
slowly varying part, due to $\hat{J}_0(r)$ and a rapidly varying
part, due to $\hat{J}_1(r)$ and the result can be a greatly reduced
magnetic shear, as seen in \Fig~\ref{fig7}.  The resulting local tearing
equation has been derived in the appendix, for the current profile of
(\ref{eq23}).  However a simpler derivation expands $q(r)$ locally around
the point, $r_1$, at the maximum of $\hat{J}(r)$, and orders the
weakened shear, $s\sim 1/m$ and $r_1^3q^{\prime \prime \prime}/q \sim
m \gg 1$. Thus:
\begin{equation}
  q(x)=q(r_1)+r_1q'\frac{x}{m}+\frac{1}{6} r_1^3 q^{\prime \prime
    \prime}\frac{x^3}{m^3}.
\label{eq24}
\end{equation}
Then, constructing $\hat{J}^\prime(x)$ from (\ref{eq23}), the high-$m$
tearing mode equation, (\ref{eq1}), can be written in the form;
\begin{align}
  \frac{d^2 \psi}{dx^2}=&\psi \left\lbrack 1 +\frac{ \kappa x
  }{x(1+\frac{1}{6} \kappa x^2)-x_s(1+\frac{1}{6} \kappa
    x_s^2)}\right\rbrack, \label{eq25}\\ 
  \kappa=&\frac{r_1^3 q^{\prime \prime
      \prime}}{m^2 r_1 q^\prime},\nonumber
\end{align}
  where $x_s$ is again the location of the resonance. We note from
   (\ref{eq24}) that a monotonic $q$ profile requires $\kappa
  >0$. \\ $~~$As in section~\ref{sec3}, stability depends on two parameters,
  $\kappa$ and $x_s$.  In order to reduce the parameter space in
  (\ref{eq25}), we have focused on three cases:
\begin{align}
(a)&\qquad  \kappa =  8 \quad & \text{monotonically increasing }q,\label{eq26}\\
(b)&\qquad  \kappa = 64 \quad & \text{with weaker shear at }x=x_s,\label{eq27}\\
(c)&\qquad  \kappa = -8 \quad & \text{non-monotonic }q(r).\label{eq28}
\end{align}
Equation~(\ref{eq25}) has then been solved to obtain values of
$\Delta(x_s)$ as the resonant location, $x_s$, is moved across the
local $q(x)$ structure.  Results for cases (a) and (b),
equations~(\ref{eq26}) and (\ref{eq27}) respectively, are shown in
\Fig~\ref{fig9}.  The solid curve corresponds to case (a) and the
dashed one to case (b). In case (c), (\ref{eq28}), we exclude
consideration of the region of triple resonance, i.e. $-1<x_s<+1$ and
\Fig~\ref{fig10} shows the value of $\Delta$ when $x_s$ falls outside
this range. Consideration of the triple resonances in case (c) raises
issues involving the different characteristic frequencies associated
with tearing at each of the three resonant surfaces, these
  frequencies being determined by diamagnetic terms and by sheared
equilibrium rotation. In addition, case (c) is likely to arise only
after an equilibrium current profile has evolved through the very
unstable weak shear scenario, case (b). It is therefore sufficient to
note that as $x_s$ approaches the location of $q_{min}$ or $q_{max}$,
the value of $\Delta$ becomes very large.

Positive values of $\Delta$, with energy available to drive
reconnection, are predicted for modes which are resonant close to, but
not at, the local maximum of the current density. Comparison of
figures~\ref{fig8} and \ref{fig9} demonstrates the validity of the
high-$m$ equation, (\ref{eq25}).
  
\section{GS2 Results}
\label{sec6}

GS2 is a radially local gyrokinetic code modeling small scale
instabilities ($k_y a\gg 1$) in a periodic flux-tube domain in
toroidal geometry. Here, $k_y=n q_r/r_r$ is the binormal wave number,
where $n$ is the toroidal mode number, and the subscript $r$ of a
quantity refers to its value at the reference radius.  The radial
variation of the metric is not retained and all plasma and magnetic
geometry parameters are linearized around their value at $r_r$. In
particular, $q\approx q_r[1+ s_r(r-r_r)/r_r]$ is used; an
approximation equivalent to (\ref{eq3}). For brevity, henceforth
we will drop the $r$ subscripts.  We use the low-flow version of GS2
\cite{lowflowgs2}, similarly as described in \cite{PIF}, and
model the current as a parallel velocity shift in the non-fluctuating
Maxwellian electron distribution. We consider a collisionless, pure
deuterium plasma where both species are gyrokinetic. We use a large
aspect ratio circular cross section geometry with no finite pressure
corrections to the flux surfaces, and we neglect compressional
magnetic perturbations.

The following parameters are used for the simulations: $u/v_i=1$,
$\beta_i=0.01$, $a/L_u=3$, $a/L_{Ti}=a/L_{Te}=a/L_n=0$,
$k_y\rho_i=0.15$, $a/R_0=0.1$, $r/a=0.5$, $s=1$. Here, $-u$ is the
electron flow speed, $v_i=(2T_i/m_i)^{1/2}$ is the ion thermal speed
with $T_i$ and $m_i$ the temperature and the mass of ions,
$\beta_i=8\pi p_i/B_0^2$ is the normalized ion pressure, $d \ln
u/dr=-1/L_u$, $d \ln n_e/dr=-1/L_n$, $d \ln T_e/dr=-1/L_{Te}$, and $d
\ln T_i/dr=-1/L_{Ti}$. Furthermore, $\rho_i=v_i/\Omega_i$ is the ion
thermal Larmor radius with the gyro-frequency $\Omega_i=e B/m_i c$,
$R_0$ is the major radius at the centroid of the flux surface. We
choose to set the temperature and density gradients to zero, thus
there are no diamagnetic corrections to the mode frequency and we can
avoid the pollution of the results with pressure gradient driven
instabilities. Therefore the current gradient is purely due to a
gradient in the electron flow speed.

We scan the safety factor $q$, which changes the current gradient
drive parameter $\sigma=-2L_s\beta_i u (ncT_i/e) (d \ln J/d \psi_0)/(
k_y^2 \rho_i^2 v_i^2)$ through the shear length $L_s=qR_0/s$. Here, we
introduced the unperturbed poloidal flux $2\pi\psi_0$. In cylindrical
geometry, this definition of $\sigma$ is equivalent to $\sigma_0$ in
(\ref{eq6}). When we set all plasma and geometry parameters to
their values specified in the previous paragraph, and let $q$ vary, we
find $\sigma=2q$. For the same set of parameters the spurious ideal
kink instabilities are found above $\sigma=2$ ($q=1$), see
\Fig~2e in \cite{PIF}. Here we will concentrate on the
region $1<\sigma<2$, where destabilization of current-gradient driven
tearing modes is possible ($\Delta>0$), as indicated by the dashed
line in \Fig~\ref{fig6}.

\begin{figure}
    \centering \includegraphics[width=0.98\columnwidth]{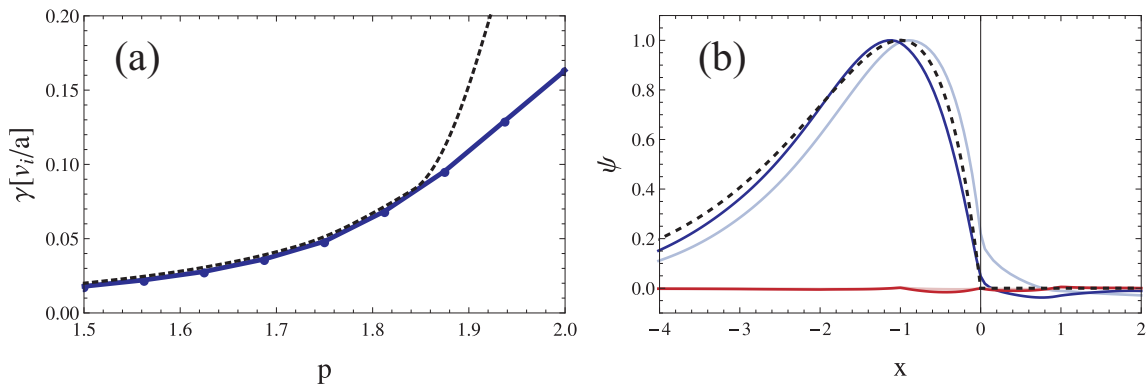}
    \caption{(a) Growth rate of the current driven instability in the
      $p$ range where the ideal kink mode is stable. Blue curve and
      symbols denote GS2 results, dotted line represents an analytical
      estimate based on \cite{drake}. (b) Typical radial structures of
      tearing modes inferred from GS2 simulations. Blue lines: real
      part, red lines (mostly overlapping, with values close to zero):
      imaginary part. Dashed line: analytical solution of marginally
      stable ideal kink). Darker curves: $p=1.9375$, lighter curves:
      $p=1.5625$.}
    \label{growthrate}
\end{figure}

Figure~\ref{growthrate}a shows the growth rate of
the tearing mode as a function of $p(\equiv\sigma)$. The symbols
represent the GS2 simulation results. The real part of
the frequency (not shown here) is very small in magnitude, consistent
with being zero within the numerical accuracy of the
calculations. We did not perform simulations below $p=1.5$
since the decreasing growth rates lead to unreasonably long simulation
times to reach convergence. The simulations barely resolve the
collisionless electron skin depth $\delta_e$ by covering an unusually
large range of ballooning angles; the number of $2\pi$ segments along
the field line is 40 in the simulations (we note that convergence of
the results with respect to resolution has been checked).

The growth rate of the collisionless tearing mode, within a constant,
is given by $\gamma\simeq\Delta_k k_y v_e/L_s$ \cite{drake}, where
$\Delta_k=\Delta'\delta_e^2$ is the width of the inner layer, with
$\delta_e^2=c^2/\omega_{pe}^2=(m_e/m_i)\rho_i^2/\beta_i$,
$\omega_{pe}$ is the electron plasma frequency, and $\Delta'$ is the
jump in $d \ln \psi/d r$ across the inner layer. Introducing the
dimensionless $\Delta=\Delta'/k_y$, we find that $\gamma
[v_i/a]\simeq\Delta
(k_y\rho_i)^2\beta_i^{-1}(sa)/(qR)[(m_eT_e)/(m_iT_i)]^{1/2}$.  We may
use the following rational approximation to describe the
$p$-dependence of $\Delta$ found from local ideal MHD calculations:
$\Delta\approx 4(p-1)/(2-p)$, shown as the dotted curve in
  \Fig~\ref{fig6}. The dotted line in \Fig~\ref{growthrate}a
represents the growth rate as estimated by the above expressions for
$\gamma$ and $\Delta(p)$. Approximations in the model break down close
to the ideal MHD instability limit, $p=2$, where $\gamma$ and
$\Delta_k$ diverge.
 
Typical radial mode structures are shown in \Fig~\ref{growthrate}b;
the $p$ values shown here are $1.5625$ and $1.9375$.  Since GS2 solves
the problem in ballooning angle $\theta$ and not in $x=(r-r_s)k_y$,
the plotted functions are obtained from the appropriate Fourier
transform of $\delta A_\|(\theta)$. The eigenfunctions $\psi$ are
normalized so that their value is $1$ at the maximum of $|\psi|$
appearing close to $x=-1$; then the blue (red) curves represent the
real (imaginary) part of $\psi$.  The eigenfunctions do not change
appreciably as $p$ is varied. In fact they very much resemble the well
known marginally stable ideal MHD result which is of the form $x e^x$
for $x<0$ and $0$ for $x>0$ (indicated with black dotted curve in the
figure).

\begin{figure}
    \centering \includegraphics[width=8cm]{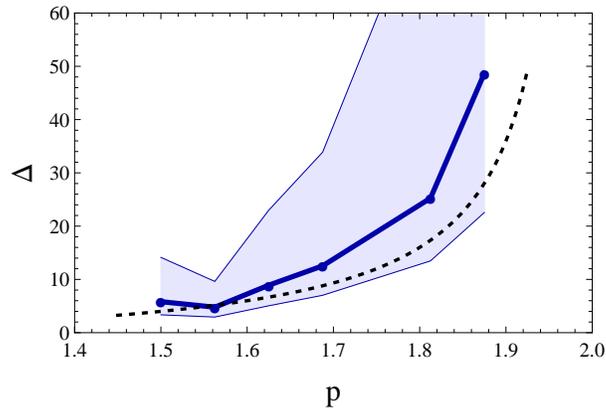}
    \caption{$\Delta$ as a function of $p$. Solid line with circle
      markers: estimated value from GS2 simulations. Light blue shaded
      area: Uncertainty in GS2 results. Dashed line: rational
      approximation of the local ideal MHD result.  }
    \label{deltaprime}
\end{figure}

Taking these $\psi(x)$ eigenfunctions, we can estimate $\Delta$ from
the GS2 simulations. As also seen in \Fig~\ref{growthrate}b, the
eigenfunctions are affected by numerical error. In the calculation of
$\Delta'$ a division by $\psi(0)$ needs to be made, which amplifies
small errors as $\psi(0)$ approaches $0$, which happens close to
the spurious ideal stability limit $p=2$. Therefore, we are unable to
determine $\Delta$ quantitatively. Figure~\ref{deltaprime} shows the
estimated values of $\Delta$ from GS2 simulations for a range of $p$
values (circle symbols and solid curve). The confidence intervals of
the results are indicated with the shaded area, obtained by perturbing
the eigenfunction within numerical uncertainties. As $p$ approaches
$2$, the uncertainties diverge; accordingly, we do not show values of
$\Delta$ above $p=1.875$. As a reference, we show the rational
approximation of the ideal MHD result by the dotted line (this is
  the same as the dotted curve of \Fig~\ref{fig6}).

\section{Summary and conclusions.}
\label{sec7}
        
This investigation was stimulated, in part, by simulation results from
GS2~\cite{PIF}, with a possible interpretation of an observed
instability as an ideal current driven kink. Such a ``ballooning
  space'' calculation assumes the neglect of $q^{\prime \prime}$ and
all higher derivatives: i.e. it is equivalent to the approximations
which lead to (\ref{eq5}) in configuration space, an equation
which, incorrectly, predicts ideal instability for values of
$\sigma_0>2$, where $\sigma_0$ is related to the ratio of current
gradient to magnetic shear.  It therefore appears that the "ideal"
instability seen in GS2 is spurious. However, the calculations
presented here show that the consequence of correctly retaining the
full functional dependence of $\sigma(x)$ in, for example,
(\ref{eq21}), is to exclude the possibility of ideal kink
instability, while still permitting unstable values of the tearing
index, $\Delta^{\prime}$, when $\sigma(0)$ exceeds a value around
unity. Hence the mode identification in terms of a "tearing/kink"
drive looks entirely justified, but its identification as an "ideal
kink" as proposed by Kadomtsev and Pogutse~\cite{KP}, should be
modified, regarding it rather as a high-$m$ tearing mode, driven
unstable by a large value of $\Delta^{\prime}$, with collisionless
reconnection provided by electron physics in the resonant layer around
$r_s$. As a consequence, it will be inappropriate to run GS2 with
values of $\sigma_0>2$. It is also of interest to note that Hegna and
Callen~\cite{HC} used a modification of (\ref{eq5}) to describe
general geometry, and as a simple way to derive a convenient formula
for $\Delta^\prime$ in toroidal systems. Although this may give a good
approximation for values of $\sigma_0<1$, where $\Delta^\prime$ is
negative, it overestimates the instability drive for $\sigma_0 > 1$
and predicts ideal instabilities for $\sigma_0>2$, where none exist.
        
We have also investigated a situation with a non-monotonic profile of
the current density, $\hat{J}(r)$, as might result from bootstrap
currents near a tokamak pedestal. Stability of high-$m$ modes is again
governed by a local equation, (\ref{eq25}), depending on two
parameters, $\kappa$ and $x_s$, where $\kappa$ is a measure of the
ratio $r_s^2 q'''/m^2 q'$ and $x_s$ is the location of the resonance
relative to the point of maximum $J$. As for the previous case, ideal
instability does not occur for any values of the $\kappa$ and $x_s$
parameters, but positive values of the tearing index, $\Delta^\prime$
can be found for sufficiently large values of $\kappa$. Such values of
$\kappa$ can arise from the low shear resulting from the near
cancellation of contributions from the background current and the
current spike.  These situations can, typically, lead to $\Delta \sim
\mathcal{O}(1)$, implying $r_s \Delta^\prime \sim \mathcal{O}(m) \gg
1$.  Since such a current spike can result from the bootstrap current
occurring naturally in the pedestal region of a tokamak H-mode plasma,
these observations have possible relevance for the interpretation of
ELMs in terms of surface ``peeling'' modes associated with tearing
modes resonant within the pedestal.  They also suggest the possibility
of influencing ELM behaviour by driving reverse currents within the
pedestal region. We note that the potentially large values of
$\Delta^\prime$ could overcome stabilising effects, such as the
Glasser effect in a torus \cite{AHG} arising from a pressure
  gradient at the resonant surface in resistive MHD, or from
  diamagnetic effects associated with the steep gradients in the
  pedestal in hotter plasmas \cite{DAHG,cowley}. To ameliorate
  the deleterious effects of large ELMs on divertor target plates,
  resonant magnetic perturbations (RMPs) have been applied to produce
  magnetic islands with the intention of driving pedestal gradients
  below the MHD stability limit. The tearing stability of the
  resulting non-symmetric equilibria is beyond the scope of this work
  but it is worth noting that the amplitude of such a RMP driven
  island depends on the value of $\Delta^\prime$ calculated in this
  work \cite{errorfields}.
       
At first sight, it is perhaps surprising that such high-$m$
calculations cannot always be reduced to a purely local calculation
involving only the current gradient and magnetic shear at the rational
surface, but instead requires that the complete structure of the $q$
profile be taken into account, albeit in a narrow region for a sharply
localised gradient in the current profile. Consideration of this
  problem is beyond the scope of local gyro-kinetic codes.
Consequently the extension to a toroidal calculation must inevitably
become two-dimensional, unlike problems amenable to the ballooning
transformation as in, e.g. the local gyrokinetic code GS2.

\ack     
This work was funded by the RCUK Energy Programme [grant number
  EP/I501045], and US Department of Energy
  grant at DE-FG02-91ER-54109 at MIT. IP is supported by the
  International Postdoc grant of Vetenskapsr{\aa}det.  To obtain
further information on the data and models underlying this paper
please contact PublicationsManager@ccfe.ac.uk. 

\appendix
\section{Approximate tearing equation for non-monotonic current profiles}
\setcounter{section}{1}        
In this appendix we demonstrate that, using the current distribution
of (\ref{eq23}) with the parameter $\mu \sim
\mathcal{O}(m)$, the global kink/tearing equation, (\ref{eq1}),
can be reduced, in the limit of high-$m$, to the form of
(\ref{eq25}), and that, for the case of the $m=4$ mode
investigated in \Fig~\ref{fig8}, the equivalent $\kappa$ value is
$51.1$.
      
 We first introduce the notation,
\begin{equation}
      p_1=\frac{\mu}{m},
\label{A1}
\end{equation}
and treat $p_1$ as $\mathcal{O}(1)$ parameter. Then, in leading order of an
expansion in $1/m$,
\begin{equation}
      r\hat{J}^\prime=-2p_1 J_1 x,
\label{A2}
\end{equation}
now with $x=m(r-r_1)/r_1$.
We next construct an expression for $1/q(r)-1/q(r_1)$ appearing in the
denominator of the current drive term of (\ref{eq1}), noting that the
contributions of the inductive current, $\hat{J}_0(r)$, and the
localised current spike, $\hat{J}_1(r)$, are simply additive, so that:
\begin{eqnarray}
      \frac{1}{q(r)}-\frac{1}{q(r_1)}&=&\frac{1}{r^2} \int_{r_1}^r s\,
      ds \lbrack \hat{J}_0(s)+\hat{J}_1(s) \rbrack \nonumber
      \\ &=&\frac{J_0 \hat{r}_1^2 x}{m} \left\lbrack -\frac{3}{2}+2
      \hat{r}_1^2-\frac{3}{4} \hat{r}_1^4 \right \rbrack +\frac{J_1
        x}{m}\left\lbrack 1-\frac{p_1 x^2}{3 m}\right\rbrack,
\label{A3}      
\end{eqnarray}
where $\hat{r}_1\equiv r_1/a$. In equation~(\ref{A3}) we retained an
$\mathcal{O}(1/m)$ correction because the leading order term is small
at low shear due to near cancellation of the contributions from $J_0$
and $J_1$ to $\mathcal{O}(1/m)$.  Finally, transforming the radial
variable in (\ref{eq1}) to $x$, the kink/tearing equation takes the
form,
\begin{equation} \frac{d^2 \psi}{dx^2}-\psi \left\{ 1
    -\frac{2p_1 J_1 x}{mx\left[J_1+J_0 \hat{r}_1^2\left(-3/2+2
        \hat{r}_1^2-(3/4) \hat{r}_1^4\right)-J_1p_1x^2/(3m)\right]}
    \right\}=0,
\label{A4}
\end{equation}
which is precisely of the same form as (\ref{eq25}) when $x_s=0$, with the
parameter, $\kappa$ given by:
\begin{equation}
      \kappa=-\frac{2p_1 J_1}{m\lbrack J_1+J_0 \hat{r}_1^2 
(-3/2+2 \hat{r}_1^2-(3/4) \hat{r}_1^4)\rbrack};
\label{A5}
\end{equation}
the above is easily generalised for non zero values of $x_s$. Note
that the above mentioned cancellation of terms in $J_0$ and $J_1$
means that $\kappa$ is formally $\mathcal{O}(1)$ but can become very
large and even change sign if the shear at $r_1$ reverses. For the
parameters of \Fig~\ref{fig7}, $J_0=2.5$, $J_1=1$, $\mu=64$,
$\hat{r}_1=0.8$, and for $m=4$, the equivalent value of $\kappa$ is
$51.1$.

\end{document}